\documentclass[aps,prd,tighten,showpacs,a4]{revtex4}

\usepackage{graphicx}
\usepackage{bm}
\usepackage{amsmath,amssymb}
\usepackage{latexsym}

\newcommand{\rp}{r_{+}}

\newcommand{\rc}{r_{c}}
\newcommand{\rcc}{\tilde{r}_{c}}

\begin{document}

\title{Lukewarm black holes in quadratic gravity.}

\author{Jerzy Matyjasek and Katarzyna Zwierzchowska}
\affiliation{Institute of Physics,
Maria Curie-
Sk\l odowska University\\
pl. Marii Curie-
Sk\l odowskiej 1,
20-031 Lublin, Poland}

\date{\today}

\begin{abstract}
Perturbative solutions to the fourth-order gravity describing
spherically-symmetric, static and electrically charged black
hole in an asymptotically de Sitter universe is constructed and discussed.
Special emphasis is put on the lukewarm configurations, in which
the temperature of the event horizon equals the temperature of 
the cosmological horizon. 
\end{abstract}


\pacs{04.50.-h, 04.70.-s,04.70.Dy}

\maketitle
\section{Introduction}

A complete  description of the
gravitational phenomena should be given by the quantum gravity, which, in turn,
may be  a part of the even more fundamental theory. Unfortunately, at the present
stage we have no clear idea how such a theory should be constructed.
It is expected, however, that
the action functional describing its low-energy approximation should
consist of the higher order terms constructed from the curvature and its
covariant derivatives to some required order. Such generalizations of the 
Einstein-Hilbert action functional necessarily  introduce a number of 
coupling constants, which, at the present state of affairs, 
should be determined empirically.

For traces of possible breakdown of the classical General
Relativity one, quite naturally, turns to cosmology. Indeed, it is possible  
that such concepts as  dark matter or/and dark energy proposed in order to
explain observational data should be abandoned in favour of modifications of the
classical gravitational Lagrangian~(see 
e.g. Ref. 1 and references cited therein). 

Among various modifications of the general relativity proposed so far,
a prominent role is played by the quadratic gravity. Various aspects
of such theories have been discussed extensively in the literature. (See for 
example Refs. 2-12).
As is well-known the motivations
for introducing linear combination of $R^{2},$ $R_{ab} R^{a b}$ and $R_{abcd} R^{abcd}$
into the gravitational action functional are numerous. For example, when invented, the equations
of quadratic gravity have been treated as an exact formulation of the
theory of gravitation. For historic informations and important references
the reader is referred to Ref. 13.
It may be considered, quite
naturally,  as truncation of series expansion of the action of the more
general theory. Moreover, from the point of view of the semi-classical
gravity it might be treated as some sort of a poor man's stress- energy
tensor, allowing in a relatively simple way to mimic, especially when the
application of the full stress- energy tensor would produce extremely
complicated result, the fairly more complex source term of the field
equations. This is especially true when the right-hand-side of the semiclassical
Einstein field equations is taken to be the  renormalized stress-energy
tensor of the quantized fields in a large mass limit constructed within the
Schwinger-DeWitt framework. Such  tensors comprise the linear
combination of the purely geometric terms and  the spin of the field
enters through the numeric coefficients~\cite{Jirinek00,jirinek01a,JMprd2009,MPLA09,Kaska}. 
As compared with the General Relativity the Lagrangian of the
quadratic gravity in four dimensions requires two additional terms
$\alpha R_{ab}R^{ab}+\beta R^{2},$ where $\alpha $ and $\beta$ are
the coupling constants. The possible third term constructed form the
Kretschmann scalar can be relegated by means of the Gauss-Bonnet term.

In general, the equations derived form the higher-order action functional are very hard to
solve. Fortunately, since the coupling constants are expected to be  
small one can easily employ a perturbative approach to the problem 
treating the classical solution of the Einstein field equations as the zeroth-order of the
approximation. Successive perturbations are therefore solutions of the
differential equations of ascending complexity. It should be noted
that although the method is clear the calculations beyond the first-
order may by intractable. Such prerturbative approach is in concord with the
philosophy of the effective theories.

Exact and approximate solutions to the equations of the quadratic gravity have been 
studied in a number of papers. Of particular interest are the spherically-symmetric
configurations which, potentially, may describe black holes. The black hole solutions
have been studied for various sources in Refs. 19-24.
Here we shall analyze a particular class of solutions of the equations of the quadratic
gravity with the cosmological term  describing lukewarm black holes\cite{Mellor1,Mellor2,Romans}. 

For the Reissner-Nordstr\"om-de Sitter class of solutions it is possible to
find configurations in which the event and cosmological horizons have the same temperature.
These solutions are known as lukewarm black holes and the interest in them
stems from the fact that if $T_{H} = T_{c}$ it is possible to construct a regular thermal state 
for a two dimensional models\cite{Lizzie1}.
Moreover, analyses of the field fluctuation indicate that this is the case in four dimensions\cite{Lizzie2,Breen}. 
(It should be noted that such configurations are prohibited in the geometry 
of the Schwarzschild-deSitter black hole.)
The natural question that arises in connection with the foregoing discussion is
whether or not it is possible to construct the lukewarm black hole in the quadratic gravity.  
And although the full, detailed answer is beyond our capabilities, it is possible to provide 
an affirmative answer to the restricted problem. Indeed, since the complexity 
of the coupled equations of the quadratic gravity and electrodynamics, even in the simplest 
case of spherically-symmetric and static geometries, hinders construction of the exact solution, 
one has to refer to the analytical approximations or numerical methods. Here we shall employ 
the perturbative approach. The obtained results can also be viewed as a first step towards 
incorporation of the quantum effects into the picture. It is because the renormalized 
stress-energy tensor of the quantized massive field may be approximated by the object
constructed from the curvature tensor, its covariant derivatives and contractions.


\section{Equations of the Quadratic Gravity}

The coupled system  of the classical electrodynamics and the quadratic gravity 
is described by the action 
\begin{equation}
S=\frac{1}{16\pi G}S_{g}+S_{m}
\end{equation}
where 
\begin{equation}
S_{g}=\int \left( R + 2 \Lambda + \alpha R^{2}+\beta R_{ab}R^{ab}\right) \sqrt{-g}\,d^{4}x,
\end{equation}
and 
\begin{equation}
S_{m}=-\dfrac{1}{16\pi }\int F_{ab} F^{ab} \sqrt{-g}\,d^{4}x,
\end{equation}
where all symbols have their usual meaning. The third possible term constructed form the
Kretschmann scalar, $R_{abcd}R^{abcd}$, may by removed from the
Lagrangian with the help of the Gauss-Bonnet invariant
\begin{equation}
R_{abcd} R^{abcd} -4 R_{ab} R^{ab} + R^{2}.
\end{equation}
Of numerical 
parameters $\alpha $ and $\beta $ we assume, as usual, that they are 
small and of comparable order, otherwise they would lead to the 
observational consequences within our solar system.  Their ultimate 
values should be  determined form observations of light deflection, 
binary pulsars and cosmological data~\cite{Odylio,accioly:2001cc,Mijic:1986iv}. 
Moreover, following Ref. 19,
we shall restrict ourselves  to spacetimes 
of small curvatures, for which the conditions
\begin{equation}
|\alpha R| \ll 1, \hspace{5mm}    |\beta R_{ab}| \ll 1
\end{equation}
hold. Although the additional constraints could be obtained 
from non-tachyon conditions~\cite{Steve,Lousto2},
which, in turn, are simple consequences of demanding the
linearized equations to posses a real mass,  we shall treat 
the parameters $\alpha$ and $\beta$ as small but arbitrary. 

Differentiating functionally the action $S$ with respect to the metric
tensor one has 
\begin{equation}
L^{ab}=G^{ab} + \Lambda g_{ab} -\alpha I^{ab}-\beta J^{ab}=8\pi T^{ab},  \label{2nd_order}
\end{equation}
where 
\begin{equation}
I^{ab}=2R^{;\,ab}-2RR^{ab}+\frac{1}{2}g^{ab}\left( R^{2}-4\Box R\right)
\end{equation}
and 
\begin{equation}
J^{ab}=R^{;\,ab}-\Box R^{ab}-2R_{cd}R^{cbda}+\frac{1}{2}g^{ab}\left(
R_{cd}R^{cd}-\Box R\right) .
\end{equation}

Let us consider the spherically symmetric and static configuration described
by the line element of the form 
\begin{equation}
ds^{2}=-e^{2\psi \left( r\right)}f(r) dt^{2}+\frac{dr^{2}}{f(r) } + r^{2}d\Omega
^{2},  \label{el_gen}
\end{equation}
where
\begin{equation}
f(r) = 1 - \frac{2 M(r)}{r}.
\end{equation}
The spherical symmetry places restrictions on the components of $F_{ab}$
tensor and its only nonvanishing components compatible with the assumed
symmetry are $F_{01}$ and $F_{23}$. Simple calculations yield 
the stress-energy tensor in the form  

\begin{equation}
T_{t}^{t}=T_{r}^{r}=-\dfrac{Q^{2}+P^{2}}{8\pi r^{4}} \equiv -\frac{Z^{2}}{8\pi r^{4}}
\end{equation}
and 
\begin{equation}
T_{\theta }^{\theta }=T_{\phi }^{\phi }=\dfrac{Q^{2}+P^{2}}{8\pi r^{4} } \equiv \frac{Z^{2}}{8\pi r^{4}},
\end{equation}
where the integration constants $Q$ and $P$ are interpreted as the electric 
and magnetic charge, respectively.

\section{Perturbative Solution}
\subsection{General Case}
To simplify calculations and to keep control of the order of terms in
complicated series expansions we shall introduce a dimensionless parameter $
\varepsilon $ substituting $\alpha \rightarrow \varepsilon \alpha $ and $
\beta \rightarrow \varepsilon \beta $. We shall put $\varepsilon =1$ at the
final stage of calculations. Of functions $M\left( r\right) $ and $\psi
\left( r\right) $ we assume that they can be expanded as 
\begin{equation}
M\left( r\right) =M_{0}\left( r\right) +\varepsilon M_{1}\left( r\right) +
\mathcal{O}\left( \varepsilon ^{2}\right)  \label{Mser}
\end{equation}
and 
\begin{equation}
\psi \left( r\right) =\varepsilon \psi _{1}\left( r\right) +\mathcal{O}
\left( \varepsilon ^{2}\right) .  \label{psiser}
\end{equation}


Consider the left hand side of Eq. (\ref{2nd_order}) calculated for the line
element (\ref{el_gen}) first. Making use of the\ above expansions and
collecting the terms with the like power one obtains 
\begin{equation}
L_{t}^{t}= \Lambda -\frac{2}{r^{2}}(M_{0}^{\prime }+\varepsilon M_{1}^{\prime
}-\varepsilon S_{t}^{t}),  \label{1st}
\end{equation}
where 
\begin{align}
S_{t}^{t}& =\beta \left( \frac{2\,M_{0}^{\prime }}{r^{2}}-\frac{
8\,M_{0}\,M_{0}^{\prime }}{r^{3}}+\frac{2\,{M_{0}^{\prime }}^{2}}{r^{2}}-
\frac{2\,M_{0}^{\prime \prime }}{r}+\frac{5\,M_{0}\,M_{0}^{\prime \prime }}{
r^{2}}-\frac{M_{0}^{\prime }\,M_{0}^{\prime \prime }}{r}\right.  \notag \\
& \left. +\frac{{M_{0}^{\prime \prime }}^{2}}{2}+M_{0}^{(3)}-\frac{
M_{0}\,M_{0}^{(3)}}{r}-M_{0}^{\prime
}\,M_{0}^{(3)}+r\,M_{0}^{(4)}-2\,M_{0}\,M_{0}^{(4)}\right)  \notag \\
& -\alpha \left( \frac{24\,M_{0}\,M_{0}^{\prime }}{r^{3}}-\frac{
8\,M_{0}^{\prime }}{r^{2}}-\frac{4\,{M_{0}^{\prime }}^{2}}{r^{2}}+\frac{
8\,M_{0}^{\prime \prime }}{r}-\frac{18\,M_{0}\,M_{0}^{\prime \prime }}{r^{2}}
-{M_{0}^{\prime \prime }}^{2}\right.  \notag \\
& \left. +\frac{2\,M_{0}^{\prime }\,M_{0}^{\prime \prime }}{r}
-4\,M_{0}^{(3)}+\frac{6\,M_{0}\,M_{0}^{(3)}}{r}+2\,M_{0}^{\prime
}\,M_{0}^{(3)}-2\,r\,M_{0}^{(4)}+4\,M_{0}\,M_{0}^{(4)}\right) 
                                      \label{1sta}
\end{align}
and primes as well as $M^{(i)}_{0}$ for $i \geq 3 $ denote 
derivatives with respect to the radial coordinate. 
It should be noted that Eq.~(\ref{1st}) does not depend on the function $\psi(r).$
The zeroth-order and the first-order equations integrated with the conditions 
$M_{0}(r_{+}) = r_{+}/2$ and
$M_{1}(r_{+}) =0,$ respectively,
give
\begin{eqnarray}
f(r) &=& 1 - \frac{\rp}{r} - \frac{Z^2}{r \rp} + \frac{\Lambda \rp^3}{3 r} + 
\frac{Z^2}{r^2} - \frac{\Lambda r^2}{3}-\left( 8\,{\frac {\Lambda\,{Z}^{2}}{{r}^{2}}}
-8\,{\frac {\Lambda\,{Z}^{2}}{rr_{+}}} \right) \alpha \nonumber \\
&-& \left( {\frac {12}{5}}\,{\frac {{Z
}^{4}}{{r}^{6}}}+4\,{\frac {{Z}^{2}}{{r}^{4}}}+{\frac {{Z}^{2} r_{+}^{3}\Lambda}{{r}^{5}}}
-{\frac {\Lambda\,{Z}^{2}}{rr_{+}}}-3\,{\frac {{Z}^{2}r_{+}}{{r}^{5}}}
-3\,{\frac {{Z}^{4}}{{r}^{5}r_{+}}}+\frac{3}{5}\,{\frac {{Z}^{4}}{rr_{+}^{5}}}
-{\frac {{Z}^{2}}{rr_{+}^{3}}} \right) \beta .\nonumber \\
\label{xx}
\end{eqnarray}
It should be noted that a correct choice of $r_{+}$ in the zeroth-order
solution requires some prescience.
Indeed, for a given $Z^{2}$ and $\Lambda$ it should allow for a positive root,
say $c,$ interpreted as the cosmological horizon which satisfies  
$r_{+} \leq c.$ Remaining roots, say $a$ and $b,$ should satisfy 
$a \leq b \leq r_{+} \leq c.$ 

On the other hand, the difference between the radial and time component of the 
tensor $L_{a}^{b}$ can be easily integrated to yield 
\begin{equation}
\psi _{1}(r)\,=\,(2\alpha +\beta )M_{0}^{(3)}-{\frac{4}{r^{2}}}(3\alpha
+\beta )M_{0}^{\prime }+C_{1},  \label{pseq}
\end{equation}
where $C_{1}$ is the integration constant. 
To determine $C_{1}$ we shall adopt the natural condition~\cite{Norma} 
\begin{equation}
g_{tt}\left(r_{\infty}\right) g_{rr}\left(r_{\infty}\right)   = -1,
\end{equation}
or, equivalently,
\begin{equation}
\psi(r_{\infty}) =0,
\end{equation}
where $r_{\infty}$ is either infinity or the cosmological horizon or the
radius of a cavity in which the system is placed (the latter case is not 
considered in this paper). 
Simple integration gives
\begin{equation}
\psi_{1}(r) = \beta Z^{2} \left(\frac{1}{r^{4}} - \frac{1}{r_{\infty}^{4}} \right)
  \label{yy}
\end{equation}
and the equations (\ref{xx}) and (\ref{yy}) provide complete first-order  
solution to the problem.  In what follows, for simplicity, we shall 
choose $r_{\infty} = \infty.$  The other choices do not influence location 
of $r_{c}$  nor the  relations describing equality of temperatures 
of the cosmological and the event horizons.

The zeroth-order solution can be expressed in a more familiar form
using, for example, the Abbott-Deser mass~\cite{Deser1}. Indeed,
if the  coefficient that stands in front of $r^{-1}$ is small
it is possible to relate it  with the Abbott-Deser mass~\cite{Deser2}. 
On the other hand,  one can introduce the parameter
\begin{equation}
{\cal M} = \frac{r_{+}}{2} + \frac{Z^{2}}{2 r_{+}} - \frac{\Lambda r_{+}^{3}}{6}
\label{parameter}
\end{equation}
to get
\begin{equation}
f(r) = 1 -\frac{2 \cal{M}}{r} + \frac{Q^{2}}{r^{2}} -\frac{\Lambda r^{2}}{3}.
\end{equation}
Simple manipulations in Eq.~(\ref{parameter}) give 
\begin{equation}
1 -\frac{2 \cal{M}}{r_{+}} + \frac{Q^{2}}{r_{+}^{2}} -\frac{\Lambda r_{+}^{2}}{3} = 0
\end{equation}
and the parameter $\cal{M}$ can be referred to as the horizon defined mass.

Returning to the first order solution one concludes that the  cosmological 
horizon is given by 
\begin{equation}
r_{c} = \rcc + \varepsilon r_{c}^{(1)},
\label{cosmo_hor}
\end{equation}
where 
\begin{eqnarray}
\rc^{(1)}  & &= - \frac{4 \Lambda\,{Q}^{2} \left( \rcc
- \rp \right) }{\tilde{\kappa}_{c}\rcc^{2}\rp}\alpha +
{\frac {3 \left( \rcc^{5}-5\,\rp^{4}\rcc +4\,\rp^{5} \right) {Q}^{4}}{10 \tilde{\kappa}_{c} 
\rcc^{6}\rp^{5}}}\beta\nonumber \\
&& -{\frac { \left( -4\,\rcc\,\rp^{3}+\Lambda\,\rcc^{4}\rp^{2}-\Lambda\,\rp^{6}
+3\,\rp^{4}+\rcc^{4} \right) {Q}^{2}}{2 \tilde{\kappa}_{c}\rcc^{5}\rp^{3}}}\beta
\label{cosho}
\end{eqnarray}
and $\tilde{\kappa}_{c} = \tilde{f}'(\rcc)/2,$ provided $\rcc$ is 
the greatest positive root of $\tilde{f}(r)=0.$  The function $\tilde{f}(r)$ 
is the unperturbed part of $f(r).$
Equations (\ref{xx}), (\ref{yy}) and (\ref{cosho})  constitute the full  (first-order) 
solution solution  to the problem.

\subsection{$T_{H} = T_{C}$ case}

Now, we shall restrict ourselves to the particular class of solutions describing 
black holes for which temperature of the event horizon equals the temperature 
of the cosmological horizon. Consider the zeroth-order solution ($\varepsilon =0$) first. 
Solving the system 
\begin{equation}
\tilde{f}(\rcc) =0 \hspace{.5cm}{\rm and} \hspace{.5cm}
 \tilde{f}'(r_{p})+\tilde{f}'(\rcc) =0
\label{luke1}
\end{equation}
with respect to $\Lambda$ and $Z^{2}$ one has
\begin{equation}
\Lambda =\frac{3}{\left( r_{+} + \rcc\right)^{2}}
\hspace{.5cm}{\rm and} \hspace{.5cm}
Z^{2} = \left(\frac{r_{+} \rcc}{r_{+} + \rcc} \right)^{2}.
\end{equation}

In the $(\rp,\rcc)-$parametrization the line element assumes the form
(\ref{el_gen}) with $\psi (r) =0$ and
\begin{equation}
\tilde{f}(r) = \left( 1 -\frac{\rp \rcc}{(\rp+\rcc)r}    \right)^{2} 
-\frac{r^{2}}{(\rp + \rcc)^{2}}
\end{equation}
whereas in the $(l,z)-$parametrization one has
\begin{equation}
\tilde{f}(r) = \left( 1 -\frac{z}{r}  \right)^{2} 
-\frac{r^{2}}{l^{2}},
\end{equation}
where $z=\sqrt{Q^{2}+P^{2}}$ and $l=\sqrt{3/\Lambda}.$
For $4z < l$ one has two pair of  roots:
\begin{equation}
\rp = \frac{l}{2}\left[1 - \sqrt{1- \frac{4 z}{l}}\right]
\hspace{.5cm}{\rm and} \hspace{.5cm}
\rcc = \frac{l}{2}\left[1 + \sqrt{1- \frac{4 z}{l}}\right],
\end{equation}
which are interpreted as the event and cosmological horizons,
and
\begin{equation}
r_{--} = -\frac{l}{2}\left[1 + \sqrt{1+ \frac{4 z}{l}}\right]
\hspace{.5cm}{\rm \phantom{a}\phantom{b}\phantom{d}} \hspace{.5cm}
r_{-} =  \frac{l}{2}\left[-1 + \sqrt{1+ \frac{4 z}{l}}\right].
\end{equation}
The positive  root $r_{-}$ is interpreted as the inner horizon whereas $r_{--}$  is a 
negative root which has no physical meaning.  

By the construction the temperature of the event horizon equals that of the 
cosmological horizon
\begin{equation}
T_{H} =T_{c} = \frac{\rcc -\rp}{2\pi (\rcc + \rp)^{2}}.
\end{equation}
Such configurations are usually referred to as the 
lukewarm black holes~\cite{Mellor1,Mellor2,Romans,Lizzie2,Breen}.  From 
the point of view of the quantum field theory in curved background the lukewarm 
black holes are special. It has been shown that for the two-dimensional models 
it is possible to construct a regular thermal state~{\cite{Lizzie1}}. Moreover, 
recent calculations of the vacuum polarization indicate that it is regular on both
the event and cosmological horizons of the D=4 lukewarm RN-dS black
holes~{\cite{Lizzie2,Breen}}. 

The lukewarm configuration in the quadratic gravity  should simultaneously satisfy 
\begin{equation}
f(\rc) =0 \hspace{0.5cm} {\rm and} \hspace{.5cm} T_{H} =T_{c}.
\label{warunki}
\end{equation}
The Euclidean section is regular at $\rp$ and $\rc$ if $\tau = it$ is periodic with 
a period $2\pi/\kappa,$ where $\kappa$ is the surface gravity.  The equality of the 
temperatures is equivalent to
\begin{equation}
\left(\frac{1}{\sqrt{g_{\tau\tau} g_{rr}}}\frac{d}{dr}g_{\tau\tau}\right)_{|r=\rp} 
+ \left(\frac{1}{\sqrt{g_{\tau\tau} g_{rr}}}\frac{d}{dr}g_{\tau\tau}\right)_{|r=\rc} =0 ,
\end{equation}
and, consequently, the lukewarm configuration is described by the above equation and  
the first equation of (\ref{warunki}). Since $r_{+}$ always denotes the exact location 
of the event horizon, mathematically, we have two relations for three parameters and 
to describe the lukewarm configuration completely  Eqs.~(\ref{warunki}) should be 
supplemented by additional condition. Here, following Ref. 32,
we shall 
adopt the point of view that the cosmological constant is not a parameter of the space 
of solutions, but, rather,  the parameter in the space of theories. To construct the 
perturbed lukewarm black hole let us substitute (\ref{cosmo_hor}) and  
\begin{equation}
Z^{2} = \left(\frac{r_{+} \rcc}{r_{+} + \rcc} \right)^{2} + \varepsilon \Delta
\label{na_Z}
\end{equation}
into the line element. Solving the system (\ref{warunki}) one obtains 
\begin{equation}
\Delta = \frac {24 \rcc^{2}\,\rp^{2}\,\alpha}{ \left( \rp+\rcc \right)^{4}} 
+ \,\frac {2\rp \left( 5\,\rcc^{4}+20\,\rcc^{2}\rp^{2}+28\,\rcc^{3}\rp+6\,\rcc\,\rp^{3}
+\rp^{4} \right) \beta}{5 \, \left( \rp+\rcc \right)^{5}}
\label{na_delta}
\end{equation}
and
\begin{equation}
\rc^{(1)} ={\frac {\beta\, \left( \rcc-\rp \right)^{2}\left( \rcc^{2}
+4\,\rcc\rp+\rp^{2} \right) }{5 \rcc\, \left( \rp+\rcc \right)^{3}\rp}}.
\label{na_cosmo}
\end{equation}
The higher-order curvature corrections to the geometry of the classical lukewarm 
black holes can readily be obtained by substituting (\ref{na_delta}) and 
(\ref{na_cosmo}) into the line element~(\ref{el_gen}), and, after expansion, 
retaining  the terms linear in $\varepsilon$. 

It is of some interest to examine a few special cases of the results (\ref{na_Z}-\ref{na_cosmo}).
In  the limit $\rcc \to \infty$ the correction $\Delta$ tends to zero
and one obtains the first order corrections to the extreme Reissner-Nordstr\"om
solution with the near-horizon geometry of the type
$AdS_{2} \times S^{2}.$  If $\beta =0$ and $\alpha \neq 0$ the correction to the
cosmological horizon vanishes and
\begin{equation}
Z^{2} = \left(\frac{r_{+} \rcc}{r_{+} + \rcc} \right)^{2} + \frac {24 \rcc^{2}\,\rp^{2}\,\alpha}{ \left( \rp+\rcc \right)^{4}}. 
\end{equation}
On the other hand, if $\alpha =0$
and $\beta \neq 0$ both $\Delta$ and $\rc$ are always of the same sign as $\beta.$
Finally, if $3\alpha + \beta =0$ both  $\Delta$ and $\rc$ are always of the same sign as $-\alpha.$
Additional constraints can be obtained from the non-tachyon conditions.

\section{Final remarks}
In this paper we have constructed the first order-correction to the geometry
of the Reissner-Nordstr\"om-deSitter black holes within the framework of the quadratic
gravity. Special emphasis has been put on the lukewarm configurations which are
characterized by the equality of the temperature of the event and cosmological horizon.
Though we have studied the consequences of the inclusion of the simplest higher-order 
correction to the gravitational action our results can shed some light on the important 
issue of the semi-classical lukewarm black holes. Since the approximate renormalized 
effective action of the quantized massive fields in the large mass limit is constructed 
solely form the curvature invariants (the type of the field is encoded in the numerical coefficients)
the analysis should follow the same lines as the analysis presented here. This strongly
support the hypothesis that solutions describing the quantum-corrected lukewarm black holes    
do exist. We intend to return to this group of problems elsewhere.

\end{document}